\documentclass[12pt]{book}%%%% latex 2e
\usepackage{sussp} %%%% change directory to suit
\usepackage{graphicx}     %%%% most useful package
\usepackage{makeidx}
\begin{document}
\headings{Escape in Hill's Problem}
{Escape in Hill's Problem}
{Douglas C Heggie}
{University of Edinburgh, UK} 

\section{Introduction and Motivation}

In the 19th century the American mathematician G.W. Hill devised a
simple and useful approximation for the motion of the moon around the
earth with perturbations by the sun.  To most dynamical astronomers
``Hill's Problem''\index{Hill's problem}
still means a model for motions in the solar system in which two
nearby bodies move in nearly circular orbits about another much larger
body at a great distance.  These lectures have, however, been
motivated by a problem in stellar dynamics\index{stellar dynamics}.

Consider a star in a star cluster\index{star cluster} which is itself in orbit about a
galaxy (Figure \ref{derivation}).  The star, cluster and galaxy take the place of the moon,
earth and sun, respectively.  The potentials of the cluster and galaxy
are not those of a point mass, and the galactic orbits of the star and
cluster may be far from circular.  Nevertheless Hill's problem is a
good starting point, and it can be modified easily to accommodate the
differences.  In section 2 we outline a derivation of Hill's
equations, and in section 3 we summarise the appropriate extensions.

Stars gradually escape\index{escape} from star clusters.  
This has been expected on
theoretical grounds for many years, ever since a paper by Ambartsumian (1938).
Recently, deep observations have confirmed this (e.g. Leon et al
2000), by revealing faint streams of stars around a number of the
globular clusters of our Galaxy.

Loosely speaking we can
say that a star can only escape if its energy exceeds some critical
energy.  The energies of stars change slightly as a result
of two-body gravitational encounters within clusters, though the time
scale on which this happens (the {\sl relaxation time\index{relaxation time} scale}) is very
long, of order $10^9$yr.  But the orbital motions of stars within
clusters have much smaller time scales of order $10^6$yr, and until
recently it was thought that escaping stars would leave on a similar
time scale.  With this assumption, relaxation is the bottleneck, and
so the escape time scale (e.g. the time taken for half the stars to
escape) should vary with the relaxation time.

\begin{figure}
[htbp]
\centering
\includegraphics[width=10cm,clip,trim=0 0 0 0,angle=-90]{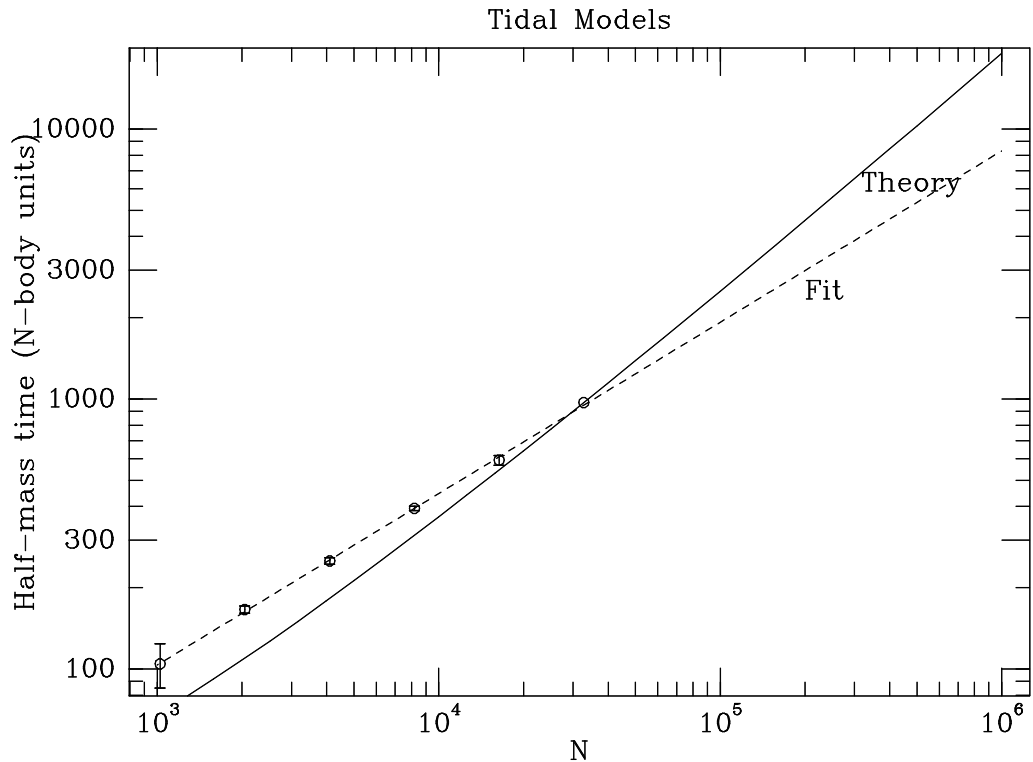}
\caption{Results of numerical experiments (Aarseth \& Heggie,
unpublished) on the escape of stars from star clusters\index{star cluster}.  The time for
half the stars to escape is plotted against the original number $N$ of stars in the
simulation.  Points are averages over several
simulations\index{$N$-body simulations} at each $N$,
except the largest value.  The continuous line shows the prediction of
theory, i.e. proportional to the relaxation time\index{relaxation time} (see text), and the dashed line is an empirical fit.}
\label{collaborative}
\end{figure}

Nowadays it is possible to simulate the evolution of modest-sized star
clusters with $3\times10^4$ or more members, and the predicted escape
time scale can be checked empirically.  Unfortunately the results
contradict the theory (Figure {\ref{collaborative}}).  As these
simulations\index{$N$-body simulations} require considerable extrapolation in particle number $N$
to be applicable to real clusters (for which $N\sim 10^6$) the error
of the theory is serious.

It turns out that the assumption of rapid escape is the main source of
error (Fukushige \& Heggie 2000, Baumgardt 2000a,b).  In fact some stars above the escape
energy never  escape\index{escape} (unless some other dynamical process comes into
play), and others take much longer to escape than had been generally
thought.

With this motivation, the remaining sections of these lectures are
devoted to the dynamics of escape.  Section 4 analyses the very
definition of escape, which is not as straightforward as in more
familiar situations.   The last two sections show some ways in
which the computation of the escape rate can be approached.  The main
result of section 5 concerns the way in which the time scale of escape depends
on the energy, and outlines how this resolves the problem of 
Figure {\ref{collaborative}}.   Much more difficult, from a theoretical
point of view, is determining the {\sl distribution} of escape times,
and some relevant ideas are introduced in Section 6.

\section{Equations of Motion}

\subsection{Derivation}

\begin{figure}
[htbp]
\centering
\includegraphics[width=10cm,clip,trim=0 0 0 0]{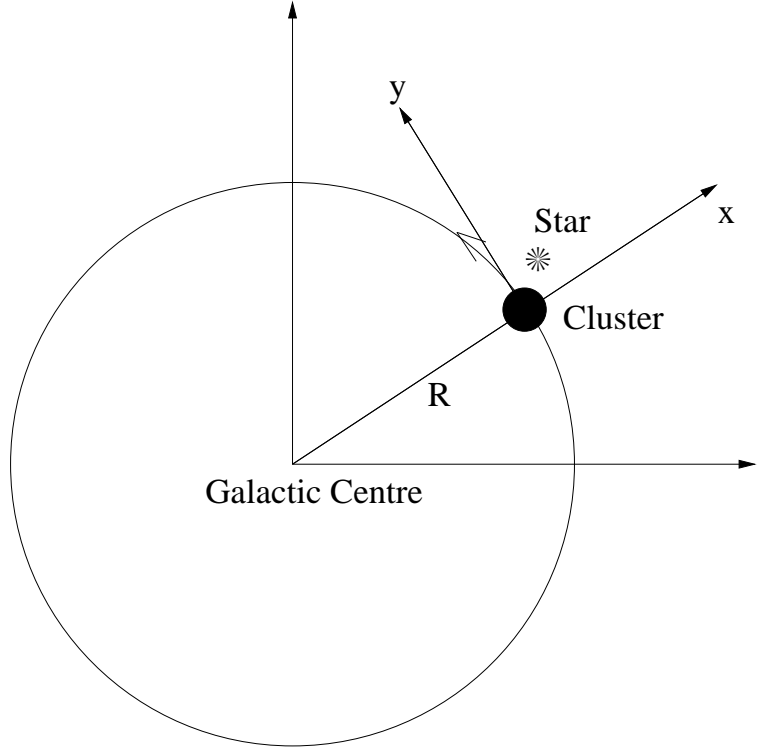}
\caption{Derivation of Hill's equations\index{Hill's equations}.  The cluster is treated as a
point mass $M_c$ in uniform circular motion of angular velocity
$\omega$ at a distance $R$ from a point-mass galaxy $M_g$.}
\label{derivation}
\end{figure}

We now outline a derivation of the equations of Hill's problem\index{Hill's problem}  in the
stellar dynamics\index{stellar dynamics} context.  To simplify matters as much as possible,
however, we treat the cluster and galaxy as point masses $M_c$ and
$M_g\gg M_c$ (Figure \ref{derivation}), and consider motion of a massless
star in the same plane of motion.

If $x,y$ are the coordinates of the star in a rotating frame centred
at the cluster centre, its velocity relative to the galaxy is $(\dot x
-\omega y, \dot y + \omega[R+x])$.  Therefore the Lagrangian\index{Lagrangian} for its
motion is ${\cal L} = \displaystyle{\frac{1}{2}\left\{(\dot x -\omega
y)^2 + (\dot y + \omega(R+x))^2\right\} + \frac{GM_g}{R^\prime} +
\frac{GM_c}{r}}$, where $r^2 = x^2 + y^2$ and $R^{\prime2} = (R+x)^2 +
y^2$.  (Note here that we are neglecting the motion of the galaxy,
which will not affect the final approximate set of equations of
motion.)

For reasons that will become clear later we switch to a canonical
formulation.  The momenta conjugate to $x$ and $y$ are
\begin{eqnarray*}
p_x &= {\cal L}_{\dot x} &= \dot x - \omega y\\
p_y &= {\cal L}_{\dot y} &= \dot y + \omega(R + x),
\end{eqnarray*}
and the Hamiltonian\index{Hamiltonian} is 
\begin{eqnarray*}
{\cal H}&=& \dot xp_x + \dot yp_y - {\cal L}\\
&=&\frac{1}{2}(p_x^2+p_y^2)+\omega(yp_x - [R+x]p_y) - \displaystyle{\frac{GM_g}{R^\prime} 
- \frac{GM_c}{r}.}
\end{eqnarray*}

The next step is common to applications in the solar system and
stellar dynamics but has a different name.  In applications to the
earth-moon-sun problem it is referred to as ``neglect of the
parallax\index{parallax}'', while in stellar dynamics it is always called a ``tidal
approximation\index{tidal
approximation}''.  (Even that phrase betrays how much the subject of stellar dynamics
owes to the celestial mechanics of the earth-moon-sun system!)  We
suppose $r\ll R$ and approximate $\displaystyle{\frac{1}{R^\prime}\simeq (1 - \frac{x}{R} +
\frac{1}{2}\frac{2x^2 - y^2}{R^2})/R}$.  We drop constant terms,
substitute $\omega^2 = GM_g/R^3$ from the equations of circular motion
(again assuming $M_c\ll M_g$), and replace $p_y\to p_y + \omega
R$. (If the other variables are not changed this transformation is
canonical.)  Then we get $\displaystyle{{\cal
H}=\frac{1}{2}(p_x^2+p_y^2)+\omega(yp_x - xp_y) -
\frac{1}{2}\omega^2(2x^2-y^2) 
- \frac{GM_c}{r}}$.

Next we write down Hamilton's equations\index{Hamilton's equations}
\begin{equation}
\dot x = {\cal H}_{p_x},
\dot p_x = - {\cal H}_x,
\end{equation}
etc.  Finally, on eliminating
$p_x$ and $p_y$, we get
\begin{eqnarray}
\ddot x - 2\omega\dot y - 3\omega^2x &= - \displaystyle{\frac{GM_c}{r^3}x}\label{unscaledx}\\
\ddot y + 2\omega\dot x &= - \displaystyle{\frac{GM_c}{r^3}y}\label{unscaledy},
\end{eqnarray}
which differ from Hill's equations\index{Hill's equations} only in notation, and then only
slightly.

\subsection{A Generalised Leapfrog}

The leapfrog\index{leapfrog algorithm} is a favourite integration algorithm for equations of
motion in stellar dynamics.  It is identical to the Verlet algorithm\index{Verlet algorithm}
of molecular dynamics.  For a one-dimensional problem with Hamiltonian
$p^2/2 + V(x)$, for example, it {\sl may} be written 
\begin{eqnarray}
x_{n+1} &=& x_n + hp_n\\
p_{n+1} &=& p_n - hV^\prime(x_{n+1}),
\end{eqnarray}
where $h$ is the time step.  Note that the new coordinate is used
immediately, which is where the algorithm differs from an Euler
algorithm\index{Euler algorithm}.  The effect is dramatic, as the long-term behaviour of the
leapfrog is {\sl much} better.  

One of the nice properties of the leapfrog is that it is {\sl
symplectic\index{symplectic}}, like a good Hamiltonian flow.  Here we show how to
construct a similar algorithm for the Hamiltonian of Hill's problem. 

Euler's algorithm would be 
\[
{\bf x}_{n+1} = {\bf x}_n + h{\cal H}_{\bf p}({\bf x_n,p_{n}}),
{\bf p}_{n+1} = {\bf p}_n - h{\cal H}_{\bf x}({\bf x_n,p_{n}}),
\]
where we have written ${\bf x} = (x,y)$ and ${\bf p} = (p_x,p_y)$.
We can make this symplectic by replacing ${\bf p}_n$ by ${\bf
p}_{n+1}$ in the arguments of the derivatives of ${\cal H}$, because
it then takes the form 
\[
{\bf x}_{n+1} = {\cal F}_{\bf p}({\bf x_n,p_{n+1}}),
{\bf p}_{n} = {\cal F}_{\bf x}({\bf x_n,p_{n+1}}),
\]
where the {\sl generating function\index{generating function}} ${\cal F} = {\bf x}_n.{\bf
p}_{n+1} + h{\cal H}({\bf x}_n, {\bf p}_{n+1})$.

Writing out these equations explicitly for the Hamiltonian\index{Hamiltonian} of Hill's
problem\index{Hill's problem}, we obtain the algorithm
\begin{eqnarray}
x_{n+1} &=& x_n + h(p_{x,n+1} + \omega y_n)\label{algorithmx}\\
y_{n+1} &=& y_n + h(p_{y,n+1} - \omega x_n)\label{algorithmy}\\
p_{x,{n}} &=& p_{x,n+1} + h(- \omega p_{y,n+1} - 2\omega^2x_n +
\frac{GM_c}{r_n^3}x_n)\label{algorithmpx}\\
p_{y,{n}} &=& p_{y,n+1} + h( \omega p_{x,n+1} +\omega^2y_n + \frac{GM_c}{r_n^3}y_n)\label{algorithmpy}.
\end{eqnarray}
These equations look horribly implicit, a common difficulty with
elementary derivations of symplectic\index{symplectic} methods, but in fact eqs.(\ref{algorithmpx}) and
(\ref{algorithmpy}) are easily solved explicitly for ${\bf p_{n+1}}$ and then
eqs.(\ref{algorithmx}) and (\ref{algorithmy}) give  ${\bf x_{n+1}}$.

\subsection{Elementary Properties}

\begin{enumerate}
\item The Hamiltonian ${\cal H}$ is time-independent, and so its value is
conserved.  Rewriting the momenta in terms of the velocity components
one finds that this value is $\displaystyle{E = \frac{1}{2}(\dot x^2 + \dot y^2) -
\frac{3}{2}\omega^2x^2 - \frac{GM_c}{r}},$
which is often referred to as the ``energy''.  Again there is another
name in the celestial mechanics community, who refer to the ``Jacobi
constant\index{Jacobi constant}'' $C = - 2E$.  In stellar dynamics this term is often applied
to $E$.  At any rate, one implication is that the motion is bounded to
the region in which $\displaystyle{-\frac{3}{2}\omega^2x^2 -
\frac{GM_c}{r}}\le E$.  The boundaries of these regions are called
{\sl Hill's curves\index{Hill's curves}} (Figure {\ref{hillcurves}}).  

It is sometimes
tempting to refer to the expression for $E$ as the Hamiltonian, and indeed the right-hand side has the same
value as ${\cal H}$.  It is, however, impossible to recover the
equations of motion from the expression for $E$.

\begin{figure}
[htbp]
\centering
\includegraphics[width=10cm,clip,trim=0 0 0 0,angle=-90]{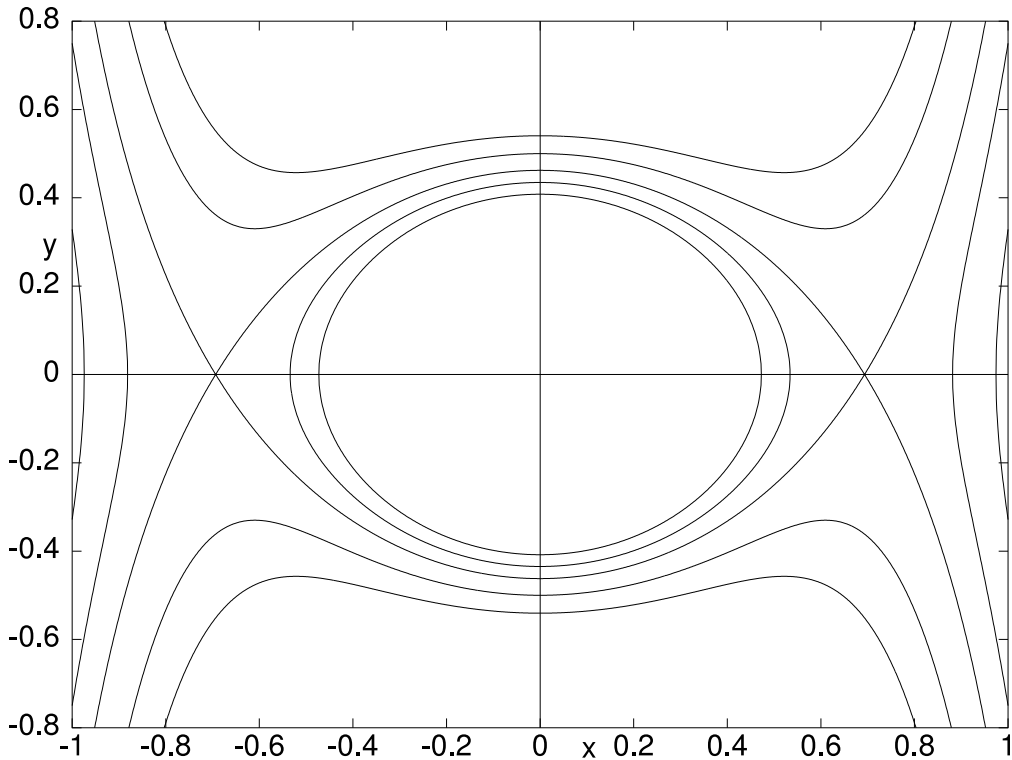}
\caption{Hill's curves.}
\label{hillcurves}
\end{figure}

\item Hill's equations\index{Hill's equations} have two equilibrium solutions, at $(x,y) = \pm
(r_t,0)$, where $r_t^3 = GM_c/(3\omega^2)$.  In stellar dynamics $r_t$
is called the {\sl tidal radius\index{tidal radius}} or {\sl Jacobi radius\index{Jacobi radius}}, and in all
subjects these points are referred to as the {\sl Lagrange points\index{Lagrange point}}
$L_1$ and $L_2$.

\item Hill's equations have an obvious symmetry: if $(x(t),y(t))$ is a
solution, then so is $(x^\prime(t),y^\prime(t)) = (x(-t), -y(-t))$.
This is quite useful for studying {\sl asymptotic} orbits\index{asymptotic orbit}.  For
example, if an orbit tends to $L_1$ as $t\to\infty$, then the orbit
obtained by this symmetry tends to $L_1$ as $t\to-\infty$.  Also, if
$\dot x(0) = y(0) = 0$, the two orbits are the same, as they satisfy
the same initial conditions.  This helps to explain the amount of
attention that has been paid in the literature to such orbits.

\end{enumerate}

\section{Variants of the Problem}

 \begin{enumerate}
\item It is not necessary that one of the bodies is massless.  Hill's
equations are also applicable to the relative motion of the moon and
earth, under solar perturbation, as in Hill's original research.  A
relatively accessible account of this research is Plummer (1918).  A
modern application is binary asteroids\index{binary asteroids} (e.g. Chauvineau \& Mignard
1990).
\item It is not necessary that the two small bodies are bound.
Another application is to near-conjunctions of coorbitals\index{coorbitals} (e.g. Murray
\& Dermott 1999).
\item When the smallest body is treated as massless, as in the star
cluster\index{star cluster} application, it is not necessary that the other bodies are
treated as  point masses.  For a spherically symmetric galaxy
potential and an arbitrary cluster potential $\phi_c$ the three-dimensional
equations of motion are
\begin{eqnarray*}
\ddot x - 2\omega\dot y + (\kappa^2- 4\omega^2)x &= -
\displaystyle{\frac{\partial\phi_c}{\partial x}}\\
\ddot y + 2\omega\dot x &= -\displaystyle{\frac{\partial\phi_c}{\partial y}}\\
\ddot z + \omega^2z &=-\displaystyle{\frac{\partial\phi_c}{\partial z}}
\end{eqnarray*}
where $\kappa$ is the {\sl epicyclic frequency\index{epicyclic frequency}} (Chandrasekhar 1942, Binney \& Tremaine
 1987) and the plane of motion of the cluster is the $x,y$ plane.  For a point-mass
 galaxy $\kappa = \omega$ and the previous equations are recovered
 (when $\phi_c = -GM_c/r$.)  

Very often the cluster potential $\phi_c$
would be chosen to be that of a {\sl King model\index{King model}} (cf. Binney \&
Tremaine 1987).  Qualitatively the most important difference from the
point-mass potential is that the depth of the potential well is
finite.  Figure {\ref{surfaceofsection}} illustrates the differences
which a change of potential can make.

\item Returning to the point mass case, we now consider the situation
in which the
motion of the cluster is elliptic, with eccentricity $e$.  There is now a formulation using the same
coordinates as in Figure {\ref{derivation}} but scaled by $R$
(so-called {\sl rotating, pulsating coordinates\index{coordinates, rotating pulsating}} $\tilde x, \tilde y$).  For coplanar motion
the equations are
\begin{eqnarray*}
\tilde x^{\prime\prime} - 2\tilde y^\prime &=& \displaystyle{\frac{1}{1+e\cos f}
( 3\tilde x - \frac{M_c}{M_g}\frac{\tilde x}{\tilde r^3})}\\
\tilde y^{\prime\prime} + 2\tilde y^\prime &=& \displaystyle{\frac{1}{1+e\cos f}
(  - \frac{M_c}{M_g}\frac{\tilde y}{\tilde r^3})},
\end{eqnarray*}
where $\prime$ denotes differentiation with respect to $f$, the true
anomaly of the cluster orbit, and $\tilde r^2 = \tilde x^2 + \tilde
y^2$.  These equations can be easily derived from the corresponding
formulation of the elliptic restricted problem\index{restricted
problem, elliptic} (Szebehely 1967).  An
important difference from the circular case is that the Hamiltonian is no
longer autonomous, and there is no Jacobi integral\index{Jacobi integral}.  
\item One can equally well treat the previous problem in rotating,
{\sl non}-pulsating coordinates with origin at the centre of the
cluster.  For coplanar motion,  a point-mass galaxy and an arbitrary
cluster potential, the equations are
\begin{eqnarray*}
\ddot x - 2\omega\dot y - (3\omega^2 - 2\displaystyle{\frac{\ddot
R}{R}})x + 2\omega\displaystyle{\frac{\dot R}{R}}y &= -
\displaystyle{\frac{\partial\phi_c}{\partial x}}\\
\ddot y + 2\omega\dot x - \displaystyle{\frac{\ddot R}{R}y} &= -\displaystyle{\frac{\partial\phi_c}{\partial y}},
\end{eqnarray*}
but the corresponding three-dimensional equations can easily be
derived for any spherical galaxy potential\index{galaxy potential} (Oh, Lin \& Aarseth 1992).
Here, of course, $\omega$ is not constant in general.
\item For a still more general galaxy potential $\phi_g$ it is simplest to use
non-rotating, non-pulsating coordinates, i.e. a coordinate frame with
origin at the cluster centre but with axes parallel to fixed
directions in space.  Then the equation of motion takes the simple
vector form $\ddot{\bf r} = -\nabla\phi_c - {\bf
r}.\nabla\nabla\phi_g$.  

Though this may well be the most useful formulation for non-circular
cluster motion, and certainly when the potential is not even spherical,
one can't help feeling that something is lost
in this simplicity.  For example, in the case of a point mass galaxy
the equation of motion is 
\begin{equation}
\ddot{\bf r} = -\nabla\phi_c - \omega^2({\bf
r} - 3({\bf r}.\hat{\bf R})\hat{\bf R})\label{vectoreom},
\end{equation}
where $\hat{\bf R}$ is the unit vector from the galaxy to the
cluster.  Now the corresponding Hamiltonian\index{Hamiltonian} is time-dependent, and it
is no longer obvious that any integral exists.  But the Jacobi
integral\index{Jacobi
integral} is still conserved, taking the form 
\[E =
\displaystyle{\frac{1}{2}\dot{\bf r}^2 - \mbox{\boldmath$\omega$}{\mathbf{.r\times\dot r}} +
\frac{1}{2}\omega^2r^2 - \frac{3}{2}\omega^2({\bf r.R})^2 + \phi_c}.
\]
This is an integral of eq.(\ref{vectoreom}), but not quite an obvious one.

 \end{enumerate}

\section{Escape Criteria\index{escape criteria}}

\subsection{Escapers}

\begin{figure}
[htbp]
\centering
\includegraphics[width=10cm,clip,trim=0 0 0 0]{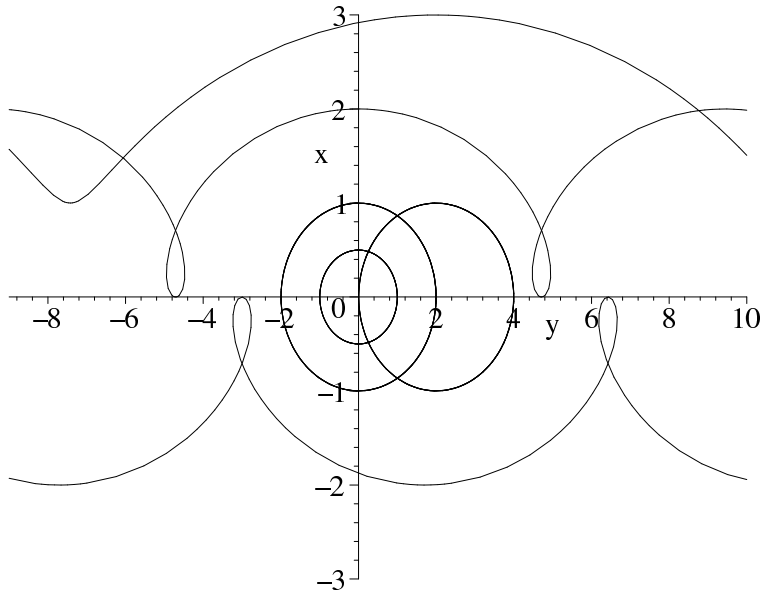}
\caption{Orbits in Hill's problem when the cluster potential is
neglected.  Note that the axes are orientated unconventionally.}
\label{epicycles}
\end{figure}

An escaping star eventually travels far from the cluster, and the
cluster potential is negligible.  If the right sides of eqs.(\ref{unscaledx}) and (\ref{unscaledy}) are
neglected we have the approximate solution 
\begin{eqnarray}
x &=& X + a\cos(t+\phi)\\
y &=& Y_0 - \displaystyle{\frac{3}{2}}Xt - 2a\sin(t+\phi),
\end{eqnarray}
where $X$, $Y_0$, $a$ and $\phi$ are constants, and we have scaled $t$
so that $\omega=1$.  Typical orbits are shown in
Figure {\ref{epicycles}}, even in places where the cluster potential
would not be negligible.

Notice that stars like to revolve or spiral anticlockwise, while the axes
are such (Figure {\ref{derivation}}) that the galaxy is far away at
the bottom and the direction of motion of the cluster is to the right.  Thus the
stellar motions are {\sl retrograde\index{retrograde motion}}.  This is typical of {\sl
epicycles\index{epicycle}}, as these motions are often called.  

Two orbits drawn in Figure {\ref{epicycles}} are centred at the location
of the cluster.  There is a family of such orbits, for varying $a$.
When the cluster potential is restored this family becomes a family of stable
retrograde satellites\index{retrograde satellite} of the cluster.  Its existence has been known for
a long time (Jackson 1913, H\'enon 1969, Benest 1971, Markellos 2000).  (In the solar system context
these are sometimes referred to as {\sl eccentric retrograde satellites},
but the reference to the heliocentric eccentricity is not illuminating
in the stellar dynamical context.)  This family, referred to as Family $f$ by
H\'enon, ranges from tiny, almost
Keplerian orbits around the origin to huge orbits far beyond the tidal
radius\index{tidal
radius}.  As a star cluster loses mass by the escape of stars, it is
conceivable that some stars in retrograde orbits are left on such
orbits well outside the tidal radius\index{tidal radius}, and it would be interesting to
look for these in $N$-body simulations\index{$N$-body simulations}.

Now consider the orbit in Figure {\ref{epicycles}} passing through the
origin.  Again such orbits of arbitrary size exist (Ross et al 1997).  Though severely
distorted by the cluster potential near the origin, they show that
stars can escape, recede to arbitrary distance, and then return to the
cluster again.  Thus distance by itself is no guarantee that escape is
permanent.   Rigorous escape criteria can be derived, but, 
to be frank, in practical terms it {\sl is} quite enough to assume
that stars that recede to a few times $r_t$ will escape; the fraction
of such stars that do not escape is tiny.

\subsection{Non-escapers}

It is easy to obtain a rigorous criterion for {\sl non}-escape, using
the simple idea behind Figure {\ref{hillcurves}}.  A particle at rest
at the $L_{1,2}$ points has energy $E_c = -
\displaystyle{\frac{3}{2}\frac{GM_c}{r_t}}$, and any star with $E\le
E_c$, and lying within the Hill curve\index{Hill's curves} passing through the Lagrangian points, can
never escape.

What now if  $E>E_c$?  We already know one set of orbits on which a
star can remain inside the cluster forever, even with energy above the
escape energy\index{escape energy}:  these are the stable retrograde satellites\index{retrograde satellite} (which move
outside the tidal radius only for energies considerably above $E_c$).

\begin{figure}
[htbp]
\centering
\includegraphics[width=10cm,clip,trim=0 0 0 0,angle=-90]{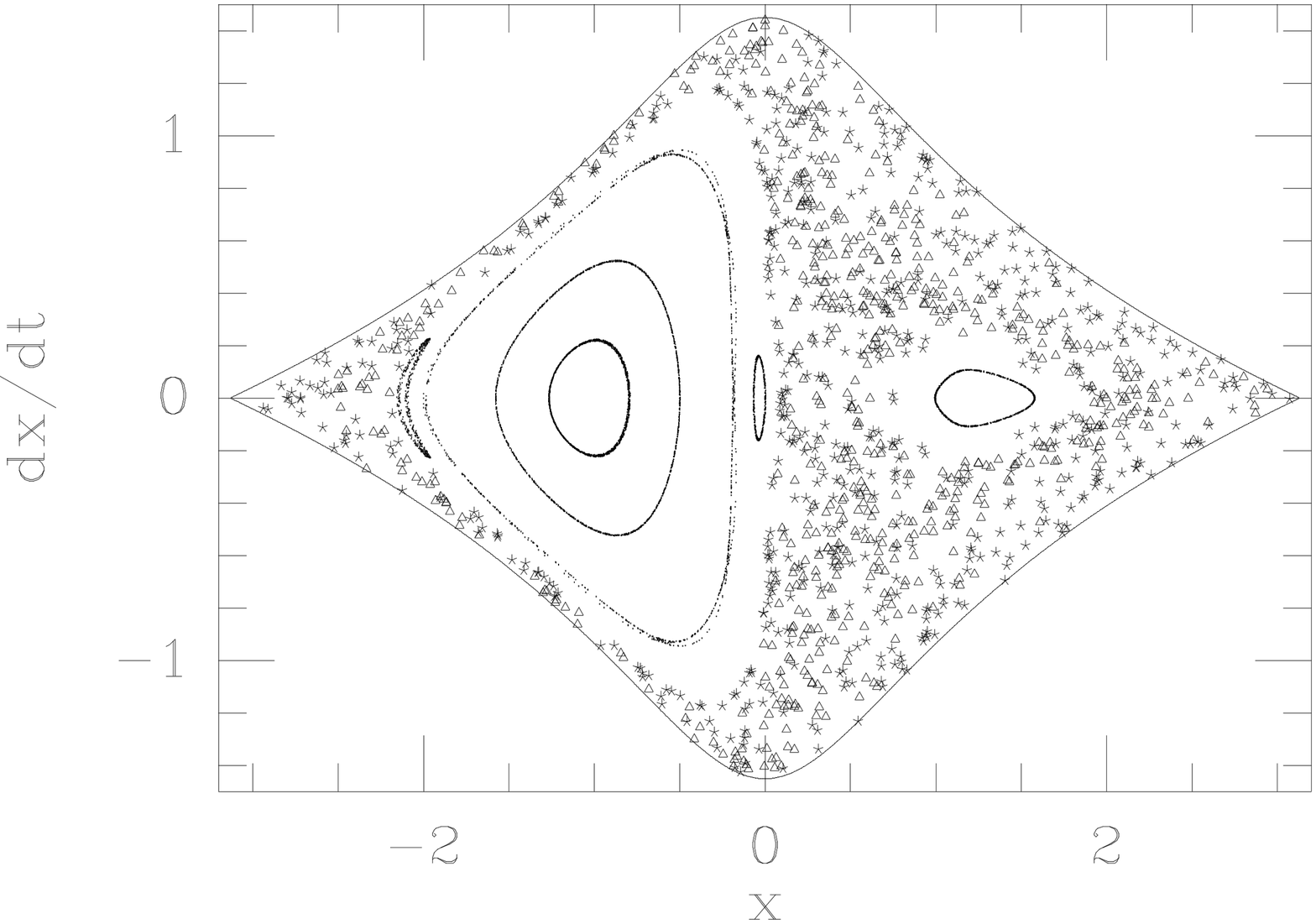}
\includegraphics[width=10cm,clip,trim=0 0 0 0,angle=-90]{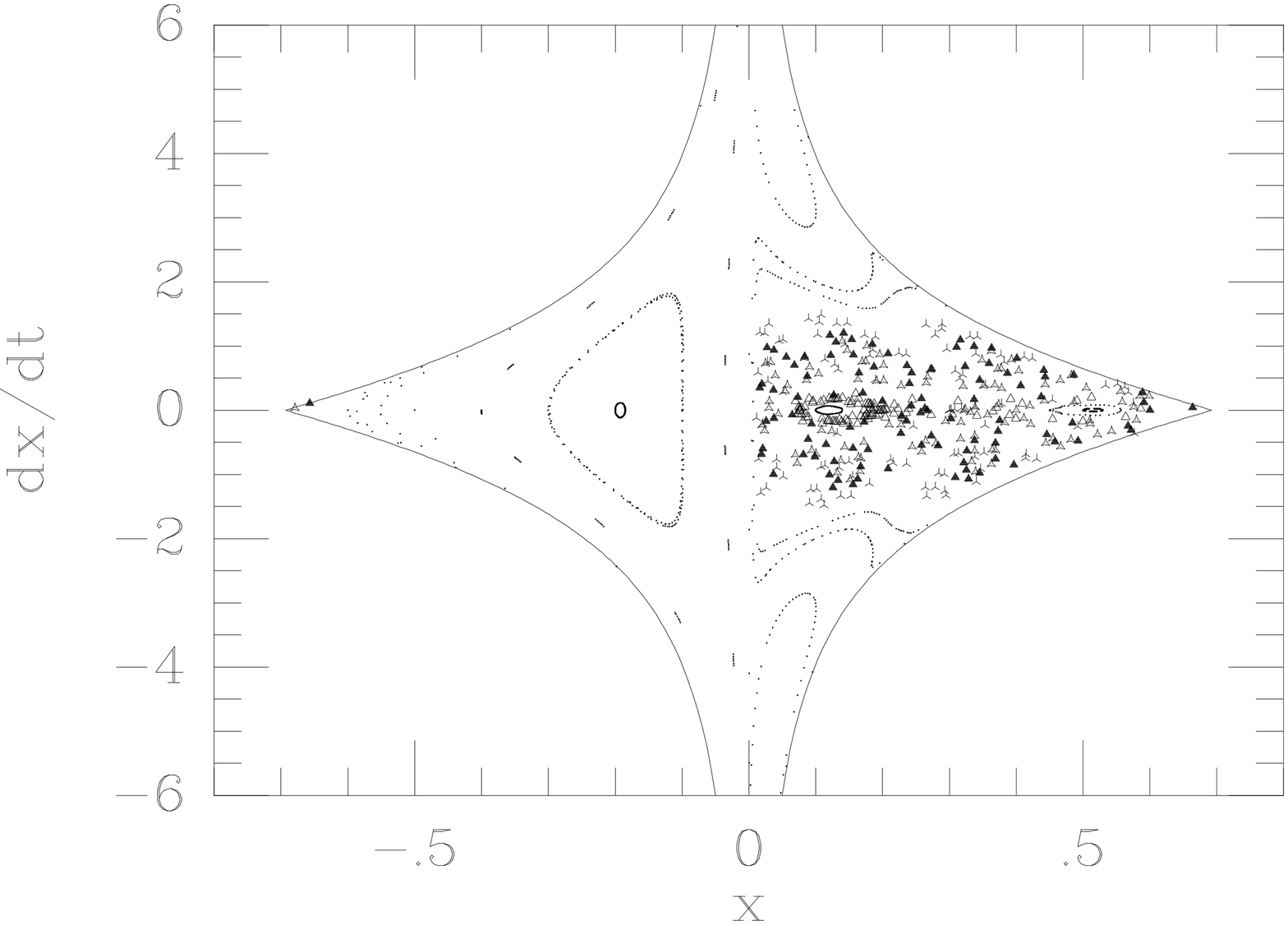}
\caption{Surface of section\index{surface of section} in the two-dimensional Hill's problem at the escape energy $E
= E_c$.  A point is generated on the surface each time an orbit
crosses the line $y = 0$ with $\dot y > 0$.
The edges of the diagram are limiting
curves\index{limiting
curve} corresponding to the condition $\dot y = 0$.  Upper diagram:
the potential 
is that of a model star cluster called a King model\index{King model} (from
Fukushige \& Heggie 2000).  Lower diagram:
point-mass potential.
}
\label{surfaceofsection}
\end{figure}

Being stable, these orbits are surrounded by a region of phase space
with the same properties.  This is illustrated by the surfaces of
section\index{surface of
section} in Figure {\ref{surfaceofsection}}.  Closed invariant curves\index{invariant curve}
predominate on the left side of the diagrams, which corresponds to
retrograde motions.  At the centre of
this nested set of curves is a fixed point corresponding to the
retrograde periodic orbit\index{retrograde periodic orbit}.  Though these diagrams are plotted for energy $E = E_c$, similar sets of invariant curves are
obtained at somewhat higher energies in the standard Hill problem (Chauvineau \& Mignard 1991,
Sim\'o \& Stuchi 2000).  They correspond to retrograde
motions of stars permanently remaining inside the cluster and with
energies above the energy of escape.  The chaotic scattering of points
on the right-hand half of the diagram would, however, generally
correspond to escaping orbits for $E>E_c$.

It is just possible that such orbits have an astrophysical relevance.
In two star clusters (Gunn \& Griffin 1979, Meylan et al 1991) there
are stars whose {\sl radial} velocity alone appears to exceed the
escape velocity.  Perhaps these are indeed stars permanently bound within the
cluster at energies above the escape energy.

\section{Escape Rate}

\subsection{Motion near the Lagrangian points\index{Lagrange point}}

Before attempting to determine the rate at which stars escape, we
study orbits in Hill's problem\index{Hill's problem} a little longer.
It is clear from the structure of Hill's curves\index{Hill's curves}
(Figure {\ref{hillcurves}}) that, at energies just above the energy of
escape, an escaper must make its way at relatively low speed through
one of the gaps in the Hill curves near $L_1$ and $L_2$.  Therefore it
pays to study motions near these points, which can be done by
linearisation of the equations of motion.

In the vicinity of $(x,y) = (r_t,0)$, when $\omega=GM_c=1$, eqs.(\ref{unscaledx}) and (\ref{unscaledy}) take the approximate
form
\begin{eqnarray*}
\ddot\xi - 2\dot\eta - 9\xi &=& 0\\
\ddot\eta + 2\dot\xi + 3\eta &=& 0,
\end{eqnarray*}
where $x = r_t + \xi$ and $y = \eta$.  These have the general solution
\begin{equation}
\left(\begin{array}{c}
\xi\\
\eta
\end{array}\right) = 
A
\left(\begin{array}{c}
-\mu\\
4-\sqrt{7}
\end{array}\right)\exp(\mu t) + 
B
\left(\begin{array}{c}
\mu\\
4-\sqrt{7}
\end{array}\right)\exp(-\mu t) + 
C
\left(\begin{array}{c}
-\nu\cos(\nu t+\theta)\\
(4+\sqrt{7})\sin(\nu t+\theta)
\end{array}\right),
\end{equation}
where $A$, $B$, $C$ and $\theta$ are arbitrary constants.
On this solution the ``energy'' is $E = E_c
\displaystyle{ + C^2(10\sqrt{7}+49) + AB(196-40\sqrt{7})}$.
Several cases have interesting properties:
\begin{enumerate}
\item $A=B=C=0$: this is the Lagrange point\index{Lagrange point},
where $E = E_{c}$ ($\equiv
\displaystyle{-\frac{3}{2}\frac{GM_c}{r_t}}$ in general).
\item $B=C=0$: this solution approaches $L_1$ as $t\to-\infty$.  It is
part of the local unstable invariant manifold\index{invariant manifold} of $L_1$, and $ E =
E_{c}$.
\item $A=C=0$: this solution approaches $L_1$ as $t\to\infty$.  It is
part of the local stable invariant manifold of $L_1$, and $ E =
E_{c}$  
\item $A = B = 0$:  the solution is periodic, and $E > E_{c}$.
Though derived in a linear approximation, there is indeed a family of
periodic solutions of the full Hill problem, parametrised by $E$
(Liapounov's Theorem\index{Liapounov's Theorem}, cf. Moser 1968).  They are named Family $a$ and
$c$ (one for each Lagrangian point) in H\'enon (1969).
\item $A=0$: part of the local stable invariant manifold of the
Liapounov orbit (Figure {\ref{orbits}}).
\item $B=0$: part of the local unstable invariant manifold of the
Liapounov orbit.
\end{enumerate}

\subsection{The flux of escapers}

\begin{figure}
[htbp]
\centering
\includegraphics[width=10cm,clip,trim=0 0 0 0,angle=-90]{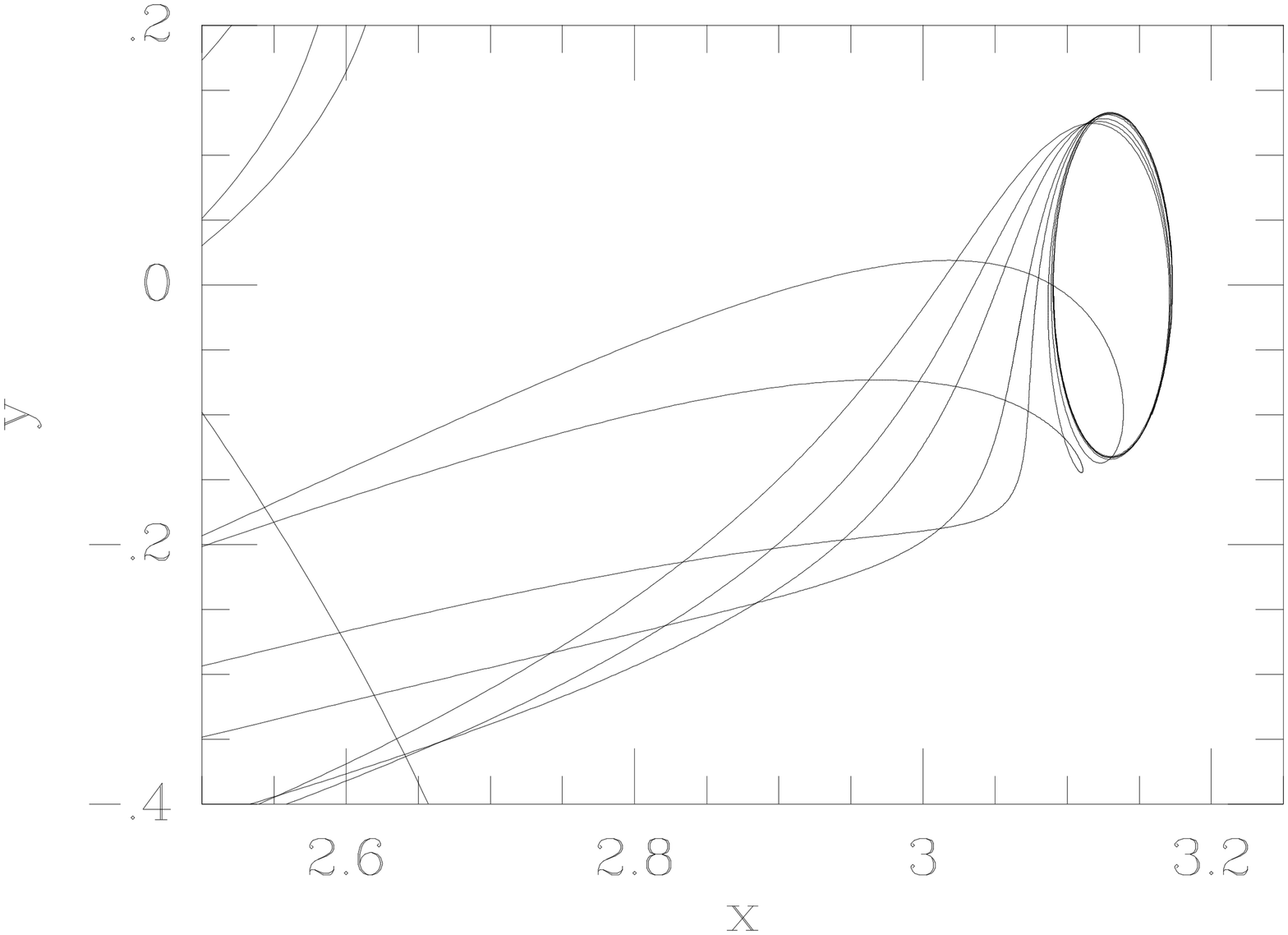}
\caption{Orbits in Hill's problem\index{Hill's problem} around one of the Lagrangian
points\index{Lagrange
point}, at a fixed energy $E$ just above $E_{c}$,   after
Fukushige \& Heggie (2000).  The potential of the cluster is that of a
King model\index{King model}.  Several orbits are shown which approach a periodic orbit\index{periodic orbit} asymptotically.  Other similar diagrams (for the point
mass potential of the usual Hill problem) are
given in Marchal (1990) and Sim\'o \& Stuchi (2000).
}
\label{orbits}
\end{figure}

Stars escaping from the interior of the star cluster have $A<0$ and
$B<0$, so that $\xi\to\pm\infty$ as $t\to\pm\infty$; thus
$C^2(10\sqrt{7}+49) < E - E_{c}$.  For fixed $E > E_{c}$, then,
this is stars ``inside'' the tube formed by the stable invariant
manifold of the Lagrange point (Figure {\ref{orbits}}).  It is quite easy to estimate the rate
at which the phase space occupied by these escapers flows out of the
cluster.  The general theory is given by MacKay (1990), though some
trivial generalisation is needed because of the Coriolis forces\index{Coriolis force} in
Hill's problem.

The rate of flow of phase space\index{phase space} (per unit energy $E_0$) is 
\begin{equation}
{\cal F} = \int_{\dot x>0, x = r_t}\dot x\delta \left(E(x,y,p_x,p_y)-E_0\right) dy dp_xdp_y,
\end{equation}
where the $\delta$-function singles out values of the
phase-space\index{phase space}
variables $x,y,p_x,p_y$ corresponding to the required energy.  This is
readily transformed to 
\begin{equation}
{\cal F} = \int_{\dot\xi>0, \xi = 0}\dot\xi\delta
\left(E(\xi,\eta,\dot\xi,\dot\eta)-E_0\right) d\eta d\dot\xi d\dot\eta
\end{equation} 
in the notation of Section 5.1.
In fact $E = \displaystyle{\frac{1}{2}\dot\xi^2 +
\frac{1}{2}\dot\eta^2 - \frac{9}{2}\xi^2 + \frac{3}{2}\eta^2 +
E_{c}}$, and so 
\begin{eqnarray*}
{\cal F} &=& \int_{\frac{1}{2}\dot\eta^2 + \frac{3}{2}\eta^2\le E_0 -
E_{c}} d\eta d\dot\eta,\\
&=& \frac{2\pi}{\sqrt{3}}(E_0 - E_{c}).
\end{eqnarray*}

This is a two-dimensional result (i.e. for the coplanar problem).  In
the three-dimensional problem it is found that ${\cal F} \propto (E_0 -
E_{c})^2$, with an equally simple coefficient.  In each case,
however, the flux\index{flux} of escaping phase space must be doubled, as there
two Lagrangian points.  

In order to turn the flux into a time scale for escape, it is only
necessary to estimate the volume of phase space inside the cluster at
energy $E$.  In turn this is given by an integral of the form 
${\cal V} = \int_{r<r_t}\delta(E - E_0) dx dy dp_x dp_y$ in two dimensions.  This does
not change much with $E_0$ in the vicinity of the  critical energy,
and there it is easily seen to be $2\pi$ times the area inside the
last closed Hill curve\index{Hill's curves} (Figure {\ref{hillcurves}}).  

It follows that the time for escape varies as $(E - E_{c})^{-2}$ in
the three-dimensional  problem, though this concerns the dimensionless
case in which $\omega = 1$.  When dimensional factors are reinserted
it turns out that the result is a time scale proportional to 
$\displaystyle{\frac{E_{c}^2}{(E - E_{c})^{2}}\frac{1}{\omega}}$.
This is a central result of these lectures.

\subsection{Numerical Methods}

It is not hard to obtain the rate of escape numerically.  One possible
procedure is the following.
\begin{enumerate}
\item Choose some $E>E_{c}$;
\item Select initial conditions at energy $E$ inside the sphere $r =
r_t$, according to some distribution (cf. Fukushige \& Heggie 2000);
\item Determine the escape time\index{escape time} $t_e$, defined to be the first time
when $r > r_t$ ({\sl pac\'e} the problem mentioned in Section 4.1);
\item Repeat 2--3 many times;
\item Compute $P(t)$, defined to be the fraction of cases with $t_e>t$.
\end{enumerate}

The third item in this procedure requires choice of a numerical
integration scheme.  Many are available, but it is worth mentioning
here one of the favourites in this subject, which is a fourth-order
Hermite scheme\index{Hermite algorithm} (cf. Makino \& Aarseth 1992).  It is a self-starting
scheme, and we illustrate it for the one-dimensional equation of
motion $\ddot x = a(x)$.  
Suppose position and velocity are known at the beginning of a
time step of length $\Delta t$, and have values $x_0$, $v_0$,
respectively.  From the equation of motion compute the initial
acceleration {\sl and its initial rate of change}, i.e. $a_0$ and $\dot
a_0$, respectively.  Compute the predicted position and velocity at
the end of the time step by
\begin{eqnarray*}
x_p &=& x_0 + v_0\Delta t + a_0\frac{\Delta t^2}{2} + \dot
a_0\frac{\Delta t^3}{6}\\
v_p &=&  v_0 + a_0\Delta t + \dot a_0\frac{\Delta t^2}{2}.
\end{eqnarray*}
Now compute the acceleration and its derivative at the end of the time
step, using $x_p$ and $v_p$.  If the results are denoted by $a_1$ and
$\dot a_1$, respectively, the values of $x$ and $v$ at the end of the
time step are recomputed by 
\begin{eqnarray*}
x_1 &=& x_0 + \frac{\Delta t}{2}(v_0 + v_1) - \frac{\Delta t^2}{12}(a_1-a_0)\\
v_1 &=& v_0 + \frac{\Delta t}{2}(a_0 + a_1) - \frac{\Delta
t^2}{12}(\dot a_1-\dot a_0).
\end{eqnarray*}

\begin{figure}
[htbp]
\centering
\includegraphics[width=10cm,clip,trim=0 0 0 0,angle=-90]{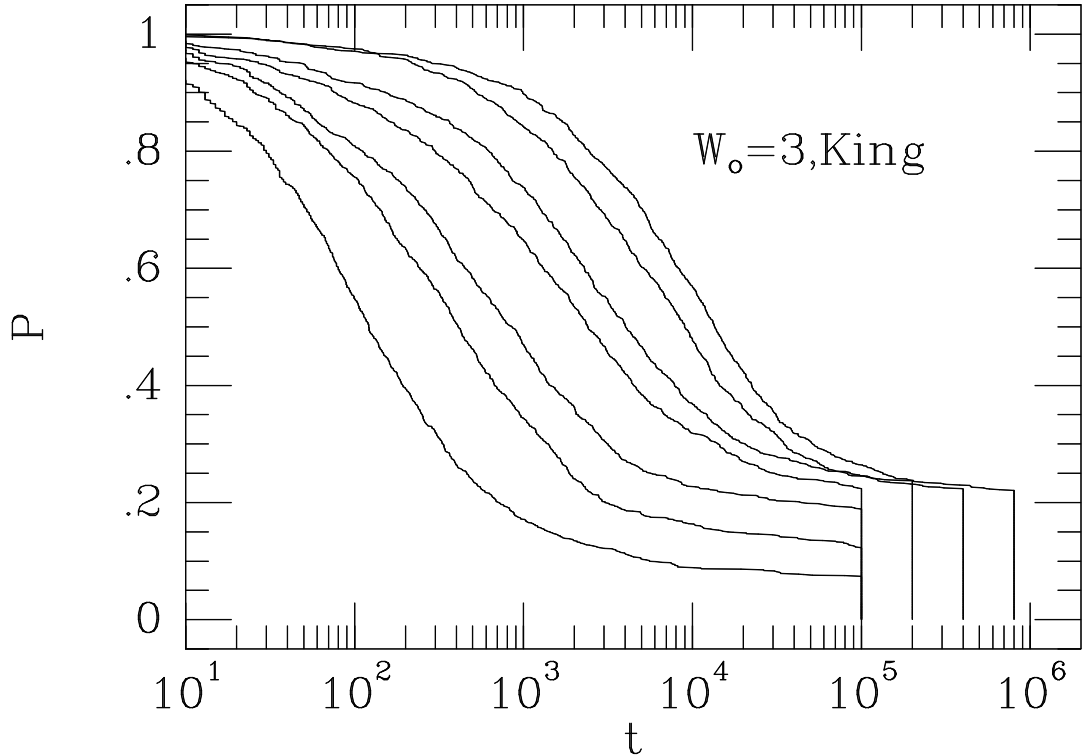}
\caption{Distribution of escape times from a generalised Hill's
problem\index{Hill's
problem}, for various values of the energy, after Fukushige \& Heggie
(2000).  }
\label{escapedistribution}
\end{figure}

Now we return to the numerical problem of determining the escape rate.
A  typical set of results is shown in
Figure {\ref{escapedistribution}}.  Curves at larger $t$ correspond to
smaller values of $E-E_{c}$.  It can be seen that these have a
horizontal asymptote well above the $t$-axis;  in other words, there
is a substantial fraction of stars for which the escape time is
extremely long.  This is not unexpected, because of the stable
retrograde motions\index{retrograde motion} shown in Figure {\ref{surfaceofsection}}.  The
fraction of such stars decreases as $E$ increases.  We also see, as
expected from Section 5.2, that the escape times\index{escape time} decrease as $E$
increases.  Indeed, if we redefine $P(t)$ to be the fraction of {\sl escapers}
with escape times $t_e > t$ (i.e. we exclude stars which never
escape), and if we rescale the values of $t$ by $(E-E_{c})^2$, the
resulting curves lie very nearly together, independent of energy (Fukushige \& Heggie 2000).

\subsection{Relaxation\index{relaxation} and Escape}

In this subsection we now show how the escape rate which we have just
determined leads to a resolution of the problems with the scaling of
$N$-body simulations\index{$N$-body simulations}, with which these lectures were motivated
(Figure {\ref{collaborative}}).  The ideas are based on those given in Baumgardt (2000b).

We imagine that stars are present in a cluster with a distribution of
energies $n(E)$.  This distribution evolves as a result of two
processes: (i) relaxation, which is a kind of diffusion process with a
characteristic time scale $t_r$; and (ii) escape, which takes place on
a time scale of order
$P\displaystyle{\left(\frac{E_{c}}{E-E_{c}}\right)^2}$, where $P$
is the orbital period of the cluster round the galaxy.

As a very simple model for this problem we shall consider the toy
model defined by the 
differential equation
\begin{equation}
\frac{\partial n}{\partial t} =
\frac{E_{c}^2}{t_r}\frac{\partial^2n}{\partial E^2} - H(E - E_{c})
\frac{1}{P}\left(\frac{E-E_{c}}{E_{c}}\right)^2n,
\end{equation}
where $n(E)dE$ is the number of stars with energies in the range $(E,
E+dE)$, and the Heaviside (unit step) function $H$ confines escape to
energies above $E_{c}$. 

There are several details missing from this problem.  First, in
addition to the diffusive term (i.e. the first term on the right side)
one needs a ``drift'' term corresponding to dynamical friction\index{dynamical friction}
(cf. Spitzer 1987, for this and other issues in what follows).  We
have also neglected the fact that the coefficient of the diffusion
term depends on $E$ and $n(E)$ in a complicated
way.  Next, one needs to take into account the effect on the energies
of the stars of the slowly changing gravitational potential of the
cluster.  Finally, we need to take into account the stars above the
escape energy whose escape time scale appears to be infinite.  If all
those factors were included, we would obtain something close to a full
Fokker-Planck equation\index{Fokker-Planck equation} for the evolution of the distribution function
in the presence of energy-dependent escape.  We shall see, however,
that this toy model is quite illuminating.

 Let us now scale $t$ by $t_r$ and let $x =
(E - E_{c})/\vert E_{c}\vert$.  Then the escape boundary occurs
at $x = 0$, and the equation transforms to 
\begin{equation}
\frac{\partial n}{\partial t} = \frac{\partial^2n}{\partial x^2} -
\alpha H(x)x^2n\label{scaledpde},
\end{equation}
where $\alpha = t_r/P$.  Now in astrophysical applications $P$ varies
with the crossing time\index{crossing time} scale in a star cluster, and so $\alpha$ varies
nearly as $N/\log N$, where $N$ is the total number of stars 
(cf. Spitzer 1987).  Therefore $\alpha$ can be used nearly as a proxy
for $N$.  

In order to estimate an escape rate we adopt the strategy pioneered by
Chandrasekhar in this context (Chandrasekhar 1943), which is to look
for a separable solution\index{separable solution} $n(x,t) = \exp(-\lambda t)y(x)$, where we
expect $\lambda>0$.  If we
impose a no-flux boundary condition at $x = -1$ (say) and the
condition that $y(x)\to0$ as $x\to\infty$, then we find that 
\begin{eqnarray}
y(x) = \left\{
\begin{array}{ll}
A\cos\left\{\sqrt{\lambda}(x+1)\right\} & \mbox{if $x<0$}\\
B\displaystyle{\int_{-\infty}^{-\sqrt{\alpha}}}
e^{\displaystyle{sx^2/2}}\vert\sqrt{\alpha}-s\vert^{\displaystyle{\lambda/4\sqrt{\alpha}
- 3/4}}
\vert\sqrt{\alpha}+s\vert^{\displaystyle{-\lambda/4\sqrt{\alpha}
- 3/4}}ds & \mbox{if $x>0$},
\end{array}\right.
\label{twosolutions}\end{eqnarray}
where $A$, $B$ are constants.  

While the first of these solutions is elementary, the second deserves
some explanation.  As Maple\index{Maple} shows, the solution of the differential
equation for $y(x)$ when $x>0$
can be written in terms of Whittaker functions\index{Whittaker functions}, and a search through
Abramowitz \& Stegun (1965) shows that these can be expressed as
integrals.  It is easier, however, to proceed directly, though the appropriate
methods are not in common use (cf. Burkill 1962).  In this particular
case, for a reason that will become clear, we first change the
independent variable to $z = x^2/2$.  Then the differential equation
becomes 
\begin{equation}
2z\ddot y + \dot y + (\lambda - 2\alpha z)y = 0\label{zde},
\end{equation}
where a dot
denotes a $z$-derivative.  Now we get down to business.
Motivated by the
inversion integral for Laplace transforms\index{Laplace transform}, we seek a solution in the
form $y(z) = \displaystyle{\int_C \exp(sz)f(s)ds}$, where both the
function $f$ and the contour\index{contour} $C$ are to be chosen.  Substituting into
eq.(\ref{zde}), we find that we require 
\[
\int_C \exp(sz)f(s)(2zs^2+s+\lambda-2\alpha z)ds = 0.
\]
No non-trivial choice of $f$ will make the integrand vanish.  We can,
however, integrate by parts to remove the $z$-dependent part of the
last factor of the integrand.  It follows that we require
\[
[2\exp(sz)f(s)(s^2-\alpha)] + \int_C\exp(sz)\left(f(s)(s+\lambda) -
\frac{d}{ds}\left\{2f(s)(s^2-\alpha)\right\}\right)ds = 0,
\]
where the first term is the end-point contribution.  Now the integral
can be made to vanish by making the integrand vanish, which in turn
requires the solution of a separable first-order differential
equation.  (Without the precaution of changing from $x$ to $z$, this
would have been a second-order equation.)  This gives the integrand in eq.(\ref{twosolutions}b).  To make the end-point
contribution vanish, we note that we require a function $y(z)$
vanishing as $z\to\infty$, and the exponential factor in the
integrand has this behaviour if we restrict the contour to $\Re s
<0$.  One obvious choice for end-point is $s = -\infty$.  For the
other we choose the negative root of $f(s)(s^2-\alpha)$, i.e. $s =
-\sqrt{\alpha}$, which works if $\lambda < 4\sqrt{\alpha}$.  In fact
the more stringent condition is the integrability of $f(s)$ at this
point, which requires $\lambda < \sqrt{\alpha}$.

Now we must 
match $y$ and $y^\prime$ at $x = 0$.  Evaluating the integral at $x=0$
is straightforward, and the transformation $s = -\sqrt{\alpha}(1+2t)$
gives a standard integral for a beta function\index{beta function}.  In order to evaluate the
derivative $y^\prime(0+)$  one cannot simply differentiate the
integral and substitute $x = 0$.  For one thing the resulting integral
diverges as $s\to-\infty$.  This behaviour is killed by the
exponential if $x$ is small and positive, and in this case one can
approximate the other factors in the integrand by their asymptotic
form as $s\to-\infty$.  Again one obtains a standard integral, this
time for a gamma function\index{gamma function}.

In the end one finds that the relation
between $\lambda$ and $\alpha$ is
\begin{equation}
\tan\sqrt{\lambda} =
\sqrt{\frac{4\sqrt{\alpha}}{\lambda}}\frac{\Gamma\left(\displaystyle{-\frac{\lambda}{4\sqrt{\alpha}}
+ \frac{3}{4}}\right)}{\Gamma\left(\displaystyle{-\frac{\lambda}{4\sqrt{\alpha}}
+ \frac{1}{4}}\right)}\label{condition}.
\end{equation}
As $\lambda/\sqrt{\alpha}\to1-$, the gamma function in the denominator
tends to infinity, and so $\sqrt{\lambda}\to0$.  Thus there is an
asymptotic regime such that $\alpha\to 0$ and
$\lambda\simeq\sqrt{\alpha}$.  If, on the other hand,
$\lambda/\sqrt{\alpha}\to0+$, it is clear that the right hand side of
eq.(\ref{condition}) tends to infinity, and so $\lambda\to\pi^2/4$.  Numerical
study shows that there is a single solution which joins these two
asymptotic regimes.

In the
second asymptotic regime ($\alpha\to\infty$, i.e. large $N$), escape\index{escape}
is very efficient, and the time scale for loss of mass, $1/\lambda$, is
determined by relaxation.  Recalling that we have scaled time by the
relaxation time\index{relaxation time}, it follows that the time to lose half the mass, say,
varies as $t_r$.
In case $\alpha$ is small, however, escape
is the bottleneck, and the escape timescale, in units of the
relaxation time, increases as $N$ (or $\alpha$) decreases.  In fact in
this regime we expect the half mass time to vary nearly as $t_r/\sqrt{N}$.
Since $t_r$ itself varies nearly as $N$ (in the units of Figure \ref{collaborative}), we
expect a time scale varying as $t_r^{1/2}$.

These results 
correspond qualitatively to what is observed (Figure \ref{collaborative}). It should
be stressed, however, that the value of this toy model is {\sl purely}
qualitative.  When one studies simulations\index{$N$-body simulations} of the evolution of star
clusters quantitatively (Baumgardt 2000b, or those in Figure \ref{collaborative}) it is found that, 
in the case of small $N$, the actual scaling is more like $t_r^{3/4}$.

We now outline Baumgardt's argument which leads to this scaling.  We
assume that the distribution of escapers (i.e. those with $E>E_{c}$)
is nearly in equilibrium.  Then eq.(\ref{scaledpde}) shows that the width of the
distribution is $x\sim \alpha^{-1/4}$.  (This scaling can also be seen
in eq.(\ref{twosolutions}b).)  
The number of such escapers is
proportional to this width, and can be estimated to be of order
$N\alpha^{-1/4}$.   The escape time scale at this energy is of order
$1/(\alpha x^2)$, $\sim\alpha^{-1/2}$, and therefore the rate of escape is
of order $N\alpha^{1/4}$.  Thus the time scale for losing (say) half
the mass is of order $\alpha^{-1/4}$ in units of the relaxation time,
i.e. the time scale of mass loss varies almost as $t_r^{3/4}$.  

It is
the assumption that the distribution of escapers reaches equilibrium
which distinguishes this estimate from the toy model discussed
previously, but the reason for this difference is not understood.

\section{Distribution of Escape Times\index{escape time, distribution of}}
   
The results of the previous section relate to the time scale on which
stars escape, and we conclude with some discussion of the actual {\sl
distribution} of escape times.  This issue has been studied in a
fairly wide variety of problems (e.g. those discussed in the book by
Wiggins 1992, and Kandrup et al 1999).  In some problems the
distribution is found to be approximately exponential, and in others
it is better approximated by a power law.  For escape in Hill's
problem\index{Hill's
problem}, the numerical experiments summarised in Section 5.3 show that
the distribution is approximately a power law, over the range of
escape times\index{escape time} that are relevant in applications and amenable to
numerical study (Fukushige \& Heggie 2000).

In this section we shall not even come close to obtaining the
distribution of escape times numerically.  We shall, however,
introduce two tools which show us how to think about this problem.
One is a suitable theoretical framework (called {\sl turnstile
dynamics\index{turnstile
dynamics}}), and the other is a toy model (H\'enon 1988) which serves
two purposes: (i) it can be used to illustrate turnstile dynamics, and
(ii) it was inspired by Hill's problem\index{Hill's problem}.

\subsection{H\'enon's Toy Model}

\begin{figure}
[htbp]
\centering

\includegraphics[width=10cm,clip,trim=0 0 0 0,angle=0]{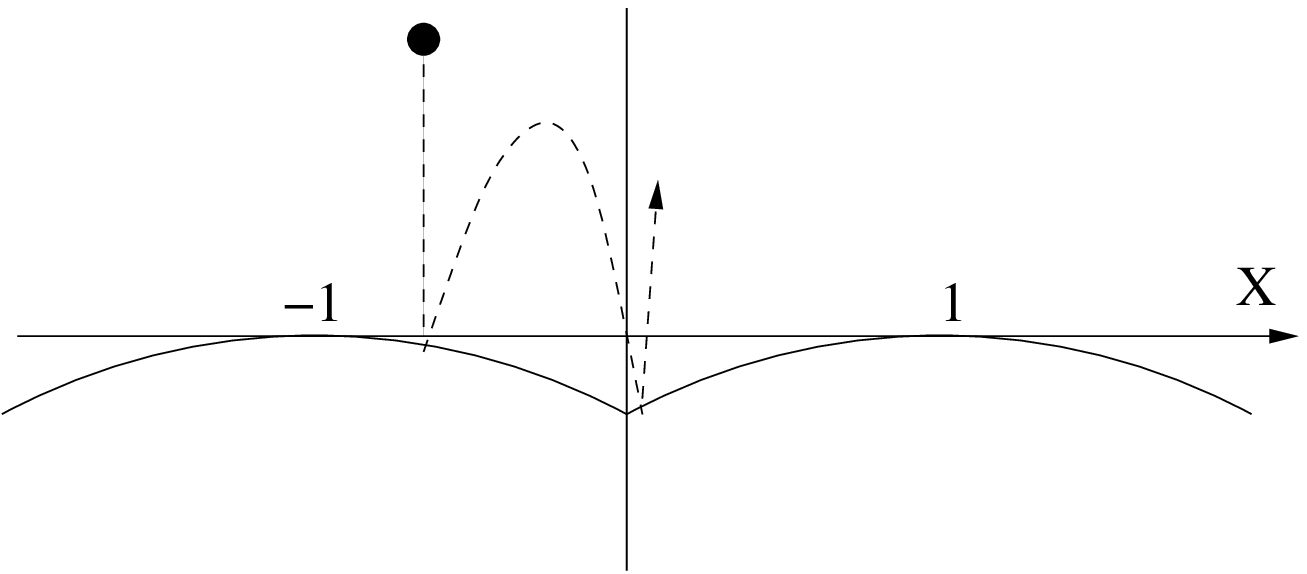}
\caption{H\'enon's billiard\index{billiard} model for Hill's problem\index{Hill's problem}.
}
\label{toymodel}
\end{figure}

We already presented a surface of section\index{surface of section} for Hill's problem, and
H\'enon's model could have been devised with the properties of the corresponding
Poincar\'e map\index{Poincar\'e map} in mind.  Physically, however, it can be thought of as
the problem of a ball falling under gravity and bouncing off two
disks (Figure \ref{toymodel}).

When the radius of the disks is very large, H\'enon showed that the
relation between conditions at each bounce takes a particularly simple
form, which is
\begin{eqnarray*}
X_{j+1} &=& X_j \cosh\psi + w_j\sinh\psi - s_j(\cosh\psi - 1)\\
w_{j+1} &=& X_j \sinh\psi + w_j\cosh\psi - (s_j\cosh\psi + s_{j+1})\tanh\frac{\psi}{2},
\end{eqnarray*}
where $\psi$ is a parameter (related to the radius of disks, the
strength of gravity, etc.), $X_j$ is the $x$-coordinate at the $j$th bounce, $w_j$ is the
tangential velocity component at this time, and $s_j =
\mbox{sign}X_j$.  (There is a tiny subtlety at $X_j=0$, which we
ignore in this exposition.)

The only non-linearities in this problem are the terms with $s$'s.
Otherwise the map is just a hyperbolic rotation\index{hyperbolic rotation} about the point $X =
\pm1$, $w = 0$, in the left and right halves of the $X,w$ plane, as
appropriate.  It is only when a point moves from one half to the other
(across the discontinuity in the surface off which the ball bounces)
that anything different happens.

\begin{figure}
[htbp]
\centering
\includegraphics[width=10cm,clip,trim=0 0 0 0,angle=0]{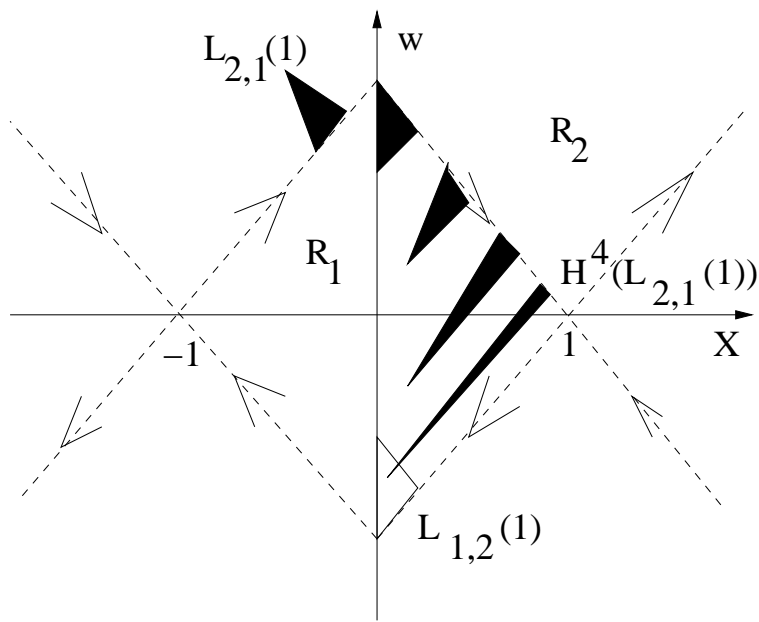}
\caption{Schematic surface of section\index{surface of section} for H\'enon's model.  The dashed lines are
the local stable and unstable invariant manifolds of the fixed points
at $(\pm 1,0)$.
}
\label{henonsection}
\end{figure}

These two points are fixed points\index{fixed point} of the map.  As usual in
such situations, a fixed point corresponds to a periodic motion, which
here refers to the ball bouncing repeatedly off either of the highest
points of the disks (Figure \ref{toymodel}).  These motions are
obviously unstable, and the fixed points on the surface of section
have local stable and unstable invariant manifolds\index{invariant manifold} which are segments
of the lines  $X = \pm 1\pm w$ (Figure \ref{henonsection}).

\begin{figure}
[htbp]
\centering
\includegraphics[width=10cm,clip,trim=0 0 0 0,angle=0]{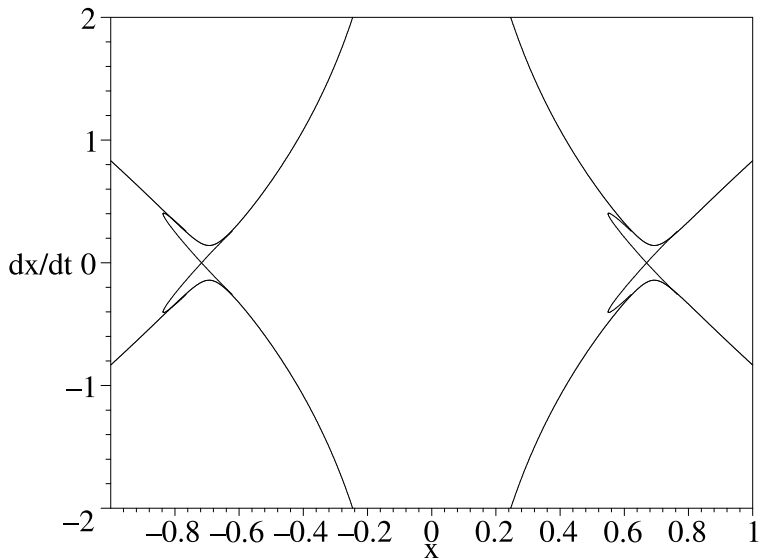}
\caption{Outline surface of section\index{surface of section} for Hill's problem\index{Hill's problem}, at some
energy $E$ just above $E_{c}$.  
The small elliptic arcs are  the local stable and unstable manifolds of the
fixed points\index{fixed point} corresponding to the Liapounov orbits, and the large
curves are the limiting curves\index{limiting curve}.
}
\label{hillsection}
\end{figure}

What has this to do with Hill's problem\index{Hill's problem}?  For one thing the unstable
periodic orbits\index{periodic orbit} have an analogy (in Hill's problem) with the Liapounov
orbits\index{Liapounov
orbit} mentioned in Sec.5.1.  Using the linearised equations derived
there it is also possible to derive equations for the local stable and
unstable invariant manifolds of the corresponding fixed points on the
surface of section (Figure \ref{hillsection}).  The main difference
between the two models is the absence, in H\'enon's model, of anything
comparable with a limiting curve.

\subsection{Turnstile Dynamics\index{turnstile dynamics}}

In Figure \ref{hillsection} it is fairly obvious how to define the
part of the section ``inside'' the cluster, and how to define the part
outside.  In Figure \ref{henonsection}, despite the absence of
limiting curves, we shall define the inside and the outside by the
naive resemblance of the two pictures.  To be more precise, the
inside ($R_1$) will be defined as the rhombus lying within the stable
manifolds of the fixed points, and the outside ($R_2$) as everything
else (Figure \ref{henonsection}).  This at least makes clear that the
boundary between the two regions is to be defined by pieces of the
stable and unstable manifolds.  This is one of the main procedural
points in the theory of {turnstile dynamics} (Wiggins 1992), which
we now  introduce via this example.  In order to apply this
theory to Hill's problem, there also we would have to define the
inside and the outside a little more carefully near the fixed points,
though we shall not dwell on the details here.

The problem of escape in H\'enon's model now focuses on the parts of the surface of
section which, under the Poincar\'e map\index{Poincar\'e map}, are exchanged between regions
$R_1$ and $R_2$.  A direct calculation (Roy 2000) shows that the
region which, on one iteration of H\'enon's map, leaves the region
$R_1$ consists of the union of two triangles.  One of these is shown
on Figure \ref{henonsection} and labelled $L_{1,2}(1)$.  The notation,
which comes from Wiggins (1992), indicates that this region is a {\sl
lobe\index{lobe}}, which moves from region $R_1$ to region $R_2$ on
{\sl one} iteration.
In the usual situation considered by Wiggins, a lobe is bounded by
parts of the stable and unstable manifolds of fixed points.  This is
only partly true in H\'enon's model.  Two parts of the boundary of the
little triangle on Figure \ref{henonsection} have this property: the
lower right, which is part of the unstable manifold of the right-hand
fixed point, and the boundary at upper right, which consists of part
of the stable manifold of the left-hand point.  The discontinuity at
$X=0$ provides the remaining part of the boundary.

We now consider {\sl capture\index{capture}} of phase space\index{phase space} from the region $R_2$
into the region $R_1$.  Again we have two triangular regions, one of
which is shown in Figure \ref{henonsection}, and labelled
$L_{2,1}(1)$, as the reader should by now appreciate.  Also shown in
the Figure are successive iterates of this lobe under the H\'enon map
$H$.  It can be seen that these remain inside $R_1$ until the map is
iterated 5 times.  The region $H^4(L_{2,1}(1))$, which is the black
triangle furthest to the lower right, intersects $L_{1,2}(1)$, and
after one further iteration this intersection leaves region $R_1$ (and
does so forever, actually.  The number of iterations that elapse
before such an intersection takes place depends on the value of
$\psi$, of course.)

Now we can see how the distribution of escape times\index{escape time,
distribution of} can be analysed,
at least in principle.  Imagine that, at $t=0$ (where $t$ counts the
number of iterations) the region $R_1$ is filled uniformly with points.  At time
$t = 1$, the area occupied by $L_{1,2}(1)$ escapes.  The same happens
at times $t = 2$, $3$, and $4$.  At time $5$, however, the number of
points that escape is given by the area of
$L_{1,2}(1)\backslash H^4(L_{2,1}(1))$.  At time $6$ the area is now 
$L_{1,2}(1)\backslash(H^4(L_{2,1}(1))\cup H^5(L_{2,1}(1)))$, and so on.

H\'enon's toy problem is unusual in that some of these calculations
can be carried out by elementary means.  In almost all problems, by
contrast, the work is necessarily numerical.  Nevertheless the ideas
of turnstile dynamics\index{turnstile dynamics} help to economise the work.  The naive way of
computing a distribution of escape times, as in Sec.5.3, is to
distribute points throughout region $R_1$ and measure how long they
take to escape.  We now see, however, that we only need to consider
the dynamics of points within $L_{1,2}(1)$ in order to reach the same
results.  This concentrates the numerical work where it is actually
needed.

When we apply these notions to Hill's problem\index{Hill's problem}, a number of additional
complicating factors arise.  In the first place the area on the
surface of section\index{surface of section} is not proportional to the volume of phase space\index{phase space}
(Binney et al 1985),
and therefore does not yield an appropriate measure of the escape
rate\index{escape
rate}.  Secondly, not all  escapers from the Hill potential actually
intersect the obvious surface of section $y = 0$.  Thirdly, the
problem is three-dimensional, and the visualisation of turnstile
dynamics becomes harder;  Wiggins' book shows some of the
complications that arise.  

On the other hand, in the planar Hill
problem some results are possible.  In particular, the analogues of
the escape and capture lobes\index{lobe}, $L_{1,2}(1)$ and $L_{2,1}(1)$, and their
iterates have been mapped out at one or two values of the energy (Roy
2000, Sim\'o \& Stuchi 2000).  For small numbers of iterations one
obtains fairly simple ovals on the surface of section.  These are the
intersections of the surface of section with the stable and unstable
invariant manifolds\index{invariant manifold} of the Liapounov orbits\index{Liapounov orbit}, i.e. structures like the
tube in Figure \ref{orbits}.  For higher numbers of iterations their
structure becomes highly convoluted, and further complicated by the
fact that, at some intersections, only part of the tube actually
intersects the surface.  

Another factor which turnstile dynamics clarifies is the relationship
between {\sl escape}, which is our interest here, and {\sl temporary capture},
which has motivated other studies (e.g. Murison 1989).

\section{Acknowledgements}

I thank J. Waldvogel, T. Quinn and C. Sim\'o for interesting
conversations about the issues in these lectures, and 
B. Chauvinea and F. Mignard for a copy of their 1991 publication.  Research with H. Baumgardt is supported by
PPARC under grant PPA/G/S/1998/00044.

\begin{susspbibliography}{99}

\bibitem{Abramowitz&Stegun1965}
Abramowitz M and Stegun I~A, 1965. \textit{Handbook of Mathematical Functions}
(New York: Dover)

\bibitem{Ambartsumian1938}
Ambartsumian V~A, 1938.  (``On the dynamics of open clusters''),
\textit{Uch. Zap. L.G.U.}, No.\textbf{22}, p.19; translated in eds
J~Goodman and P~Hut
\textit{Dynamics of Star Clusters}, Proc. IAU Symp. 113, 521--524 (Dordrecht: Reidel)

\bibitem{Baumgardt2000a}
Baumgardt H, 2000a.  In \textit{Dynamics of Star Clusters and the Milky Way},
eds S~Deiters, B~Fuchs, A~Just, R~Spurzem and R~Wielen, in press  (San
Francisco: ASP)

\bibitem{Baumgardt2000b}
Baumgardt H, 2000b. Scaling of $N$-body calculations, \textit{MNRAS}, submitted

\bibitem{Benest1971}
Benest D, 1971. 
Elliptic restricted problem for sun-jupiter:
existence of stable retrograde satellites at large distance,
\textit{A\&A} \textbf{13} 157--160

\bibitem{Binneyetal1985}
Binney J, Gerhard O~E and Hut P, 1985. Structure of surfaces of
section \textit{MNRAS} \textbf{215} 59--65

\bibitem{Binney&Tremaine1987}
Binney J and Tremaine S, 1987. \textit{Galactic Dynamics} (Princeton: Princeton
Univ. Press) 

\bibitem{Burkill1962}
Burkill J~C, 1962. \textit{The Theory of Ordinary Differential Equations}
(Edinburgh: Oliver \& Boyd)

\bibitem{Chandrasekhar1942}
Chandrasekhar S, 1942. \textit{Principles of Stellar Dynamics}
(Chicago: Univ. of
Chicago Press;  also New York: Dover, 1960)

\bibitem{Chandrasekhar1943}
Chandrasekhar S, 1943. 
Dynamical friction II. The rate of escape of
stars from clusters and the evidence for the operation of dynamical
friction, 
\textit{ApJ} \textbf{97} 263--273

\bibitem{Chauvineaumignard1990}
Chauvineau B and Mignard F, 1990.
Dynamics of binary asteroids. I -
Hill's case \textit{Icar} \textbf{83} 360--381

\bibitem{Chauvineaumignard1991}
Chauvineau B and Mignard F, 1991. \textit{Atlas of the Circular Planar Hill's
Problem}  (Grasse: Obs. de la Cote d'Azur)

\bibitem{Fukushige&Heggie2000}
Fukushige T and Heggie D~C, 2000. The time scale of escape from star
clusters \textit{MNRAS} in press

\bibitem{Gunn&Griffin1979}
Gunn J~E and Griffin R~F, 1979. Dynamical studies of globular clusters based on photoelectric radial velocities of individual stars.
 I - M3 \textit{AJ} \textbf{84} 752--773
                            
\bibitem{Henon1969}
H\'enon M, 1969.
Numerical exploration of the restricted
problem. V. Hill's case: periodic orbits and their stability
\textit{A\&A} \textbf{1}
223--238

\bibitem{Henon1988}
H\'enon M, 1988. Chaotic scattering modelled by an inclined
billiard \textit{Physica D} \textbf{33} 132--156

\bibitem{jackson1913}
Jackson J, 1913. Retrograde satellite orbits \textit{MNRAS} \textbf{74} 62--82

\bibitem{Kandrupetal1999}
Kandrup H~E, Siopis C, Contopoulos G and Dvorak R, 1999. Diffusion
and scaling in escapes from two-degree-of-freedom Hamiltonian
systems 
\textit{astro-ph}/9904046

\bibitem{Leonetal1999}
Leon S, Meylan G and Combes F, 2000. Tidal tails around 20 Galactic
globular clusters. Observational evidence for gravitational disk/bulge
shocking \textit{A\&A} \textbf{359} 907--931

\bibitem{MacKay1990}
MacKay R~S, 1990. Flux over a saddle \textit{Phys. Lett. A} \textbf{145} 425--427

\bibitem{Makino&Aarseth1992}
Makino J and Aarseth S~J, 1992. On a Hermite integrator with
Ahmad-Cohen scheme for gravitational many-body problems \textit{PASJ}
\textbf{44} 141--51

\bibitem{Marchal1990}
Marchal C, 1990. \textit{The Three-Body Problem}  (Amsterdam: Elsevier)

\bibitem{Markellos2000}
Markellos V~V, 2000. \textit{private communication}

\bibitem{Meylanetal1991}
Meylan G, Dubath P and Mayor M, 1991. Two high-velocity stars shot
out from the core of the globular cluster 47 Tucanae  \textit{ApJ} \textbf{383} 587--593

\bibitem{Moser1968}
Moser J~K, 1968. Lectures on Hamiltonian Systems \textit{Mem. AMS} \textbf{81} 1--60;
also in eds R~S MacKay and J.D. Meiss, 1987.  \textit{Hamiltonian Dynamical Systems}
(Bristol: Adam Hilger) 

\bibitem{Murison1989}
Murison M~A, 1989.  The fractal dynamics of satellite capture in the
circular restricted three-body problem {\textit AJ} {\textbf 98}, 2346--59 and 2383--6

\bibitem{Murray&Dermott1999}
Murray C~D and Dermott S~F, 1999. \textit{Solar System Dynamics}
(Cambridge:  Cambridge Univ. Press)

\bibitem{Ohetal1992}
Oh K~S, Lin D~N~C and Aarseth S~J, 1992. 
Tidal evolution of globular
clusters. I. Method
\textit{ApJ} \textbf{386} 506--18

\bibitem{Plummer1918}
Plummer H~C, 1918. \textit{An Introductory Treatise on Dynamical Astronomy}
(Cambridge:  Cambridge Univ. Press; also New York: Dover, 1960)

\bibitem{Rossetal1997}
Ross D~J, Mennim A and Heggie D~C, 1997.
Escape from a tidally
limited star cluster 
\textit{MNRAS} \textbf{284} 811--814

\bibitem{Roy2000}
Roy A, 2000.  PhD Thesis, Univ. of Edinburgh, in preparation

\bibitem{Simo&Stuchi2000}
Sim\'o C and Stuchi T~J, 2000.
Central stable/unstable manifolds and
the destruction of KAM tori in the planar Hill problem
\textit{Physica D}, \textbf{140}, 1--32

\bibitem{Spitzer1987}
Spitzer L, Jr, 1987. \textit{Dynamical Evolution of Globular
Clusters} (Princeton:  Princeton Univ. Press)

\bibitem{Szebehely1967}
Szebehely V, 1967. \textit{Theory of Orbits} (New York: Academic Press)

\bibitem{Wiggins1992}
Wiggins S, 1992. \textit{Chaotic Transport in Dynamical Systems}
(Berlin: Springer-Verlag)

\end{susspbibliography}

%\newpage\mbox{}
%\printindex

\end{document}